\documentclass{elsarticle}
\usepackage{graphicx}
\usepackage{psfrag}
\usepackage{makeidx}  
\usepackage{multirow}
\usepackage{psfrag}
\usepackage{caption}
\usepackage{amsmath}

\usepackage{amsfonts}

\begin{document}

\title{Combining neural networks and signed particles to simulate quantum systems more efficiently,\\Part III}

\author[ca]{Jean~Michel~Sellier$^*$}
\author[ca]{Ga\'{e}tan~Marceau~Caron}
\author[ca]{Jacob~Leygonie}
\address[ca]{Montr\'{e}al Institute for Learning Algorithms,\\Montr\'{e}al, Qu\'{e}bec, Canada\\$^*$\texttt{jeanmichel.sellier@gmail.com}}

\begin{abstract}
This work belongs to a series of articles which have been dedicated to the combination of
signed particles and neural networks to speed up the time-dependent simulation of quantum
systems. More specifically,   the suggested networks are utilized to compute the function
known as the Wigner kernel, a multi-dimensional function defined over the phase space. In
the first paper, we suggested a network which is completely defined analytically,   based
on prior physics knowledge, and which does not necessitate any training process. Although
very useful for practical applications,    this approach keeps the same complexity as the
more standard finite difference methods.    In facts, the number of hidden neurons in the
networks  must be equal to the number of cells in the discretized real space  in order to
provide reliable results. This technique has    the advantage of drastically reducing the
amount of memory needed though and,   therefore, represents a step ahead in the direction
of computationally less demanding simulations of relatively big  quantum systems.  In the
second work, we presented a different architecture which, instead,     has generalization
capabilities,   and therefore requires a training process.    Although more convenient in
terms of computational time,   it still uses a similar structure compared to the previous
approach,    with less neurons in the hidden layer but with the same expensive activation
functions, i.e. sine functions.  In this work, the final part of this series, we focus on
neural networks without any prior physics based knowledge.   In more details, the network
now consists of different hidden layers which are based on more common,          and less
computationally expensive, activation functions such as rectified linear units, and which
focus on predicting only one column of the discretized Wigner kernel at a time   (instead
of the whole kernel as done previosuly). It requires a training process, to determine its
weights and biases, which is performed over a dataset consisting of         a position, a
potential, and its corresponding kernel on that specified position.    This new suggested
architecture proves to accurately learn the transform from the space        of  potential
functions to the space of Wigner kernels  and performs very well during the simulation of
quantum systems. As a matter of facts,     it allows a drastic reduction of the amount of
memory required   along with     a much lower   computational burden ($~20$ times faster).
Finally, as a validation test, we simulate a simple,  but meaningful,     one-dimensional
quantum system consisting of a Gaussian wave packet representing an electron impinging on
a potential barrier, and comparisons with other time-dependent approaches    are provided.
This represents one further step ahead to achieve        fast and reliable simulations of
time-dependent quantum systems.
\end{abstract}


\maketitle

\section{Introduction}

This work represents the last part of a series of three papers         (\cite{PhysA-2017},
\cite{PhysA-2018})      which explore different ways of exploiting neural networks in the
framework of     the signed particle formulation of quantum mechanics \cite{SPF}.    This
relatively new formalism is based on the  concept  of an ensemble of field-less Newtonian
particles completely describing the system,        provided with a sign, and which create
couples of new signed particles according to some pre-computed probability.    Because of
its simplicity, and in spite of its recent appearance,   it has already been applied with
success to the simulation of a relatively big number of different situations, both in the
context of single- and many-body systems,  showing unprecedent  advantages    in terms of computational resources
\cite{PhysRep} (the interested reader can find practical               examples involving
time-dependent simulations     of quantum many-body systems in \cite{JCP-01}-\cite{IJQC}).
Even if this innovative approach has important unique features,  one of its bottleneck is
represented by the computation of the so-called Wigner kernel,        a multi-dimensional
integral which is necessary in order to evolve signed particles.  In facts, this function
is defined over a phase space which dimensions  are equal to $2 \times N \times d$, where
$N$ is the number of physical bodies involved and $d$ is  the dimension of the real space
($d=1, 2, 3$).   Therefore, it can quickly become a critical aspect for the simulation of
quantum systems since both the amount of memory and the time required for its computation
are cursed by the total dimensionality of the system or,     equivalently,         by the
dimensionality of the configuration space.

\bigskip

Recently, the use of Artificial Neural Networks (ANN) to address the problem of computing
the Wigner kernel rapidly and reliably was initially suggested   by one of the authors in
\cite{PhysA-2017}.      In this preliminary investigation, a new technique was introduced
which is based on an appropriately tailored ANN in the context     of the signed particle
formalism.   The suggested network architecture has the peculiar feature of not requiring
any training phase, since mathematical and physical knowledge of the problem is enough to
retrieve the weights and biases analytically. Although this first approach has shown some
important advantages, its computational complexity remains an issue (see \cite{PhysA-2018}
for a complete discussion on this topic).       Subsequently, a more general approach was
introduced which uses a different network architecture and,  unlike the previous approach,
requires a training process (see \cite{PhysA-2018}).   This method has the main advantage
of reducing the complexity of the ANN (since it utilizes less neurons in the hidden layer)
and, therefore, allows a faster computation of the Wigner kernel.

\bigskip

In this work,  we present a completely different approach based on more common techniques
coming from the field of machine learning. In particular,   we suggest a new architecture
consisting of three hidden layers with neurons which implement  the rectified linear unit
(ReLU) as their activations, i.e. a function which is drastically less demanding in terms
of computational       resources when compared to the previously utilized sine activation
functions. In order to show the validity of this new approach,         we apply it to the
standard problem consisting of  a one-dimensional Gaussian wave packet going towards with
a potential barrier.       For the sake of clarity and completeness, comparisons with our
previously implemented techniques are presented.

\bigskip

This paper is organized as follows. In the next section, we shortly introduce and discuss
the previous techniques which were able  to combine the use of signed particles with ANNs.
Afterwards, we proceed with the description of the new ANN architecture   which radically
improves the previous techniques. Finally, as it is usual, a validation test is performed
to assess the reliability and speed of our new suggested approach and conclusive comments
are provided.        The authors strongly believe that this series of works increases the
chances to continue to depict robust, fast and reliable ways to   simulate time-dependent
quantum systems,    with a potential impact in important fields such as quantum chemistry
and the design of quantum computing electronic devices.

\section{Neural Network Architectures}

In this section,  we start by providing the context of the problem we face   in this work.
In particular, for the sake of self-consistency,   we report the third postulate on which
the signed particle formulation of quantum mechanics is based.    Then, we proceed with a
short description of the solutions proposed in    \cite{PhysA-2017} and \cite{PhysA-2018}.
Finally,      we present a novel ANN architecture which is reliable and fast but does not
carry the computational burden of the previously suggested techniques.

\subsection{The third postulate of the signed particle formulation}

The signed particle of quantum mechanics is based on three postulates    which completely
determines the time-dependent evolution of an ensemble of signed particles and,   in turn,
of a quantum system.       In this section, we briefly discuss about  postulate II, which
eventually represents the bottleneck of this novel approach.     Postulates I and II have
been discussed elsewhere and can be summarized as {\sl{1)}} a quantum system is described
by an ensemble of signed field-less classical particles    which completely describes the
system (essentially, in the same way the wave function does),     and {\sl{2)}} particles
with opposite signs but equal position and momentum always annihilate.       Postulate II,
in full details, is reported below for a              one-dimensional, single-body system
(the generalization to many-dimensional,     many-body systems is easily derived, see for
example \cite{PhysRep}).

\bigskip

{\sl{{\bf{Postulate.}} A signed particle, evolving in a given potential $V=V \left( x \right)$, behaves as a
field-less classical point-particle which, during the time interval $dt$, creates a new pair of signed particles
with a probability $\gamma \left( x(t) \right) dt$ where
\begin{equation}
 \gamma\left( x \right) = \int_{-\infty}^{+\infty} \mathcal{D}p' V_W^+ \left( x; p' \right)
\equiv \lim_{\Delta p' \rightarrow 0^+} \sum_{M = -\infty}^{+\infty} V_W^+ \left( x; M \Delta p' \right),
\label{momentum_integral}
\end{equation}
and $V_W^+ \left( x; p \right)$ is the positive part of the quantity
\begin{equation}
	V_W \left( x; p \right) = \frac{i}{\pi \hbar^2} \int_{-\infty}^{+\infty} dx' e^{-\frac{2i}{\hbar} x' \cdot p} \left[ V \left( x+x' \right) - V \left( x-x'\right)  \right],
\label{wigner-kernel}
\end{equation}
known as the Wigner kernel \cite{Wigner}. If, at the moment of creation, the parent particle has sign $s$,
position $x$ and momentum $p$, the new particles are both located in $x$, have signs $+s$ and $-s$, and momenta $p+p'$ and $p-p'$ respectively,
with $p'$ chosen randomly according to the (normalized) probability $\frac{V_W^+ \left( x; p \right)}{\gamma(x)}$.}}

\bigskip

Therefore,              one can view the signed particle formulation as made of two parts:
{\sl{1)}} the evolution of field-less particles, which is always   performed analytically,
and {\sl{2}} the computation of the kernel (\ref{wigner-kernel}),        which is usually
performed numerically. In particular, the computation of the Wigner kernel can, sometimes,
represent a problem in terms of computational implementation     as it is equivalent to a
multi-dimensional integral which complexity increases rapidly with the dimensions  of the
configuration space.    It is clear that a naive approach to this task is not appropriate
(for more technical details the reader is encouraged to visit \cite{nano-archimedes}  for
a free implementation of the approach).

\bigskip

We now briefly describe the previously suggested methods         of \cite{PhysA-2017} and
\cite{PhysA-2018}, and then present our new approach.

\subsection{Previous approaches}

At a first glance,  it might seem relatively simple to train an ANN to predict the kernel
(\ref{wigner-kernel}) once a potential is provided.  In other words, one simply looks for
a map between the space of vectors  representing  physical  potentials  and  the space of
the corresponding kernels, a rather common problem in machine learning  (usually referred
to as supervised learning).

\bigskip

It initially appeared to the authors   that a simple naive approach based on a completely
general ANN aiming to learn the mapping by itself  would be difficult to depict and train.
Therefore, we first decided to exploit some prior knowledge           to make the problem
approachable (interestingly enough, similar conclusions have been obtained             in
\cite{Science2017} and \cite{Nature2017}).        Surprisingly it was eventually shown in
\cite{PhysA-2017},                                                that by performing some
relatively simple algebraic manipulation, it is possible to obtain such ANN {\sl{without}}
any training since    we are in front of one rare example of neural network which weights
can be found analytically.                                                            The
network consists of an input layer, a hidden layer and an output layer.   The input layer
receives a discretized position in the phase space,   indicated by the couple of integers
$(i, j)$, along with a discretized potential $V=V(x)$,          represented by the vector
$\left[ V_1 = V(x_1), \dots, V_n = V(x_n) \right]$. To speed up   the network, an initial
pre-computation of the angles $\theta_l$          and the corresponding sine functions is
performed. Eventually, these values are utilized to    define the activation functions of
the hidden layer and, consequently,  an weighted average    is computed in the last layer
which represents the output of the network (see \cite{PhysA-2017} for all details).
This quite uncommon  approach brings   two important advantages: {\sl{1)}}, it  completely
avoids  the  need  to compute the Wigner kernel everywhere on the phase-space,   {\sl{2)}},
the curse of dimensionality affecting the amount of memory required is completely removed
from the picture. One important drawback remains though since this  network still retains
the initial complexity of the problem.

\bigskip

Eventually, in order to give generalization capabilities and,    therefore,    reduce the
numerical complexity of the problem,          an improvement to the previous approach was
suggested based on introducing an arbitrary number of parameters (in other words, weights
and biases) to be learned during a training process \cite{PhysA-2018}.    To achieve such
goal,   one starts from the previous approach and carefully simplifies it to do not loose
accuracy. In particular,     one introduces the physically reasonable hypothesis that   a
given potential can be approximated by a baricentric interpolation.    Although arbitrary
and dependent on the discretization lenght in the configuration space,    this assumption
offered a first simple  way to improve   our previous approach in terms of generalization
and, therefore, numerical performance. Eventually, this interpolation is generalized to a
weighted average of the potential values and the weights are learnt by the network during
the training process.    In order to find those values,   we search for the weights which
provide the best network approximation of the function $V_W = V_W(x; p)$    (representing
the dataset) by means  of a standard machine learning method known as stochastic gradient
descent.

\subsection{On learning to compute the Wigner kernel}

In this paragraph,    we now describe a complete approach to the problem of computing the
Wigner kernel by means of      a completely general neural network (i.e. not based on any
prior physical knowledge). In more details, the purpose of this ANN is to learn a mapping
$f:\mathbb{R}^m \times I \rightarrow \mathbb{R}^n$,       which generates a column of the
discretized Wigner kernel once a discretized potential (in other words, a vector) and  an
index in $I$ are provided,  and where the index associated to the column represents     a
position in the discretized configuration space. To achieve it, we use a totally agnostic
approach to quantum physics by casting the problem    into a standard supervised learning
problem where the mapping is learned from a finite set of training examples.   By using a
simple feedforward neural network,    we show that it is possible to learn such a mapping
which, in turn,    allows the simulation of quantum systems efficiently and accurately in
the context of the signed particle formulation discussed above.

However, it is well known that, compared to approximating the kernel (\ref{wigner-kernel})
by some standard numerical method (e.g. finite differences), the price to pay consists in
not having any theoretical guarantees that   the ANN      will generalize well enough for
{\sl{any}} discretized potential in $\mathbb{R}^m$, or even worse,  for all indices for a
given potential.    Therefore, we are forced to trade-off this property for a significant
speedup in terms of computational time without losing too much accuracy. In any case,  in
the next section,  we will empirically demonstrate that the approximation error is indeed
negligible for a large family of physically meaningful potentials and, in particular, for
the barrier potential (a quintessential validation test for quantum simulations).  In the
light of these results, one can claim that   {\it{the model used does not only generalize
in the statistical sense,            it also learns a very good approximation of the real
transformation $f$}}.


\bigskip

\begin{figure}
\centering
\includegraphics[width=1\textwidth]{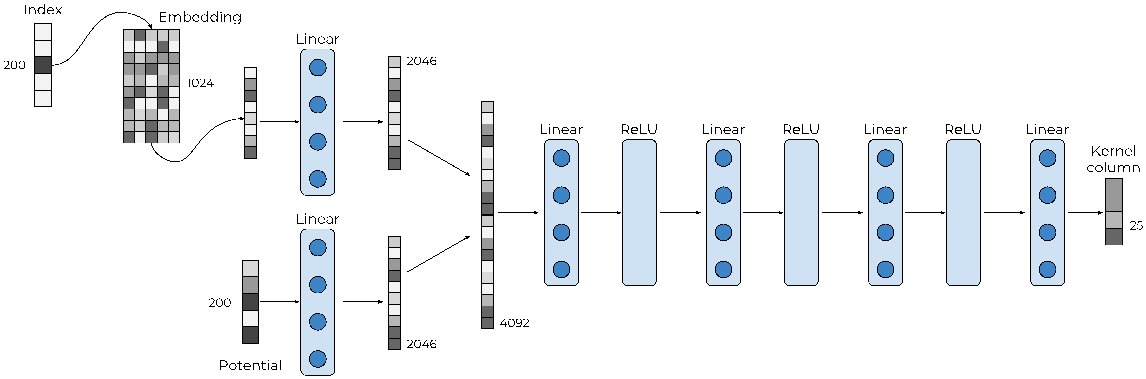}
\caption{Visual representation of the ANN suggested in this work. The index is represented
by a one-hot vector which selects one column of the embedding. This column is linearly transformed into a larger vector space and concatenated
with the potential and, successively, this new vector is transformed by a composition of linear
transformations and ReLU activations. The hidden representations after the concatenation
module all have $2046$ dimensions (not shown on the figure). The circles mean that the module contains parameters.}
\label{fig:model}
\end{figure}

In more details, to achieve the goal described above,  the main effort is represented  by
the optimization of two antagonist objectives, i.e. computational accuracy  and speed, by
selecting the appropriate ANN architecture.    Thus, to facilitate this task, we restrict
the search space of ANN architectures to feedforward neural networks.      This is mainly
motivated by their simplicity along with their ability           to easily capture global
interactions between the inputs\footnote{Interestingly, we also experimented with convolutional neural
networks, but their performances were not satisfactory  to justify their use in a quantum
simulation.}.  Moreover, for a fixed potential, the diversity of outputs becomes important
when we vary the index,     since this diversity is controlled only by changing the index
variable. To increase the influence of this input variable,     we encode it as a one-hot
vector and we embed it in a high-dimensional space with five times  more entries than the
number of positions, as depicted in Fig. \ref{fig:model}. Eventually,   this embedding is
transformed again into a hidden representation that is concatenated to         the hidden
representation of the potential. Finally, this vector is then processed   by two pairs of
rectified linear units (ReLU) followed by a linear transformation before generating   the
kernel column. At the end, this model has $19,550,233$ free parameters (for the case discussed in the next section), and yet it is the
smallest architecture which achieves accurate computations that we were able to find.     Even though the number of
parameters might sound quite large at a first glance,       we actually achieve a speedup
about $18$ times faster than the simplest finite difference quadrature       (such as a C
implementation of the midpoint and the rectangular quadrature methods).      In fact, the
architecture presented in Fig. \ref{fig:model}          can be efficiently implemented on
Graphical Processing Units (GPUs),        without any effort, with the use of modern deep
learning libraries    such as Pytorch \cite{pytorch}. Consequently, most of the computations consists of
matrix multiplications which allows   to increase   the speedup by using large batches of
data at once.      On table \ref{table:time}, we show the average time in milliseconds to
compute a whole kernel.           We take into account the initial cost associated to the
initialization of the GPU.  This cost is then amortized by computing several columns of a
kernel inside a single batch of data.

\begin{figure}
\centering
 \includegraphics[width=0.75\textwidth]{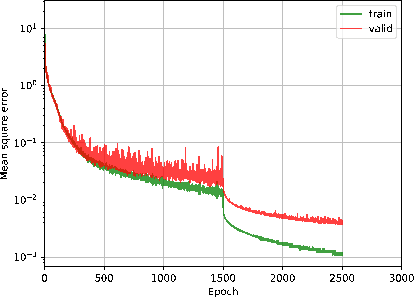}
\caption{The learning curve of the model on the training and validation sets. The MSE is given on a log-scale. After $1500$ epochs, we divide the learning rate by ten.}
\label{fig:learning}
\end{figure}

\begin{figure}
\centering
 \includegraphics[width=0.49\textwidth]{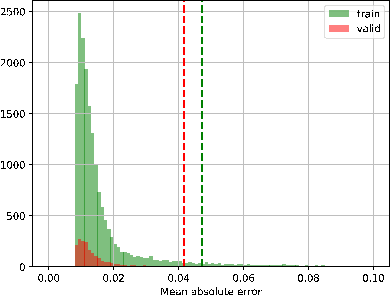}
 \includegraphics[width=0.49\textwidth]{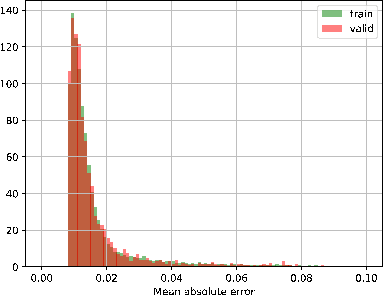}
\caption{Comparison of the mean absolute errors between the training set and the validation set in absolute (left-hand side) and relative (right-hand side) frequency.}
\label{fig:error}
\end{figure}

\begin{figure}
\centering
\includegraphics[width=0.49\textwidth]{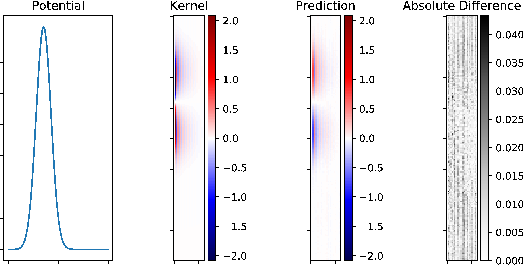}
\includegraphics[width=0.49\textwidth]{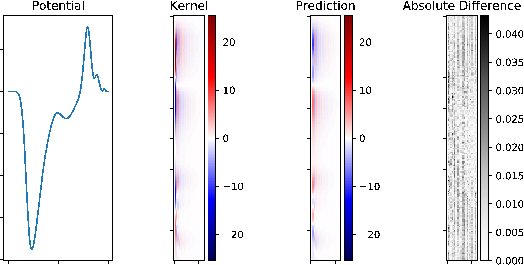}
\includegraphics[width=0.49\textwidth]{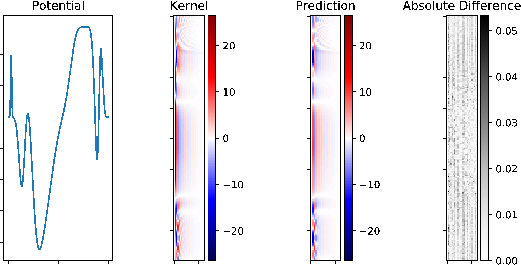}
\includegraphics[width=0.49\textwidth]{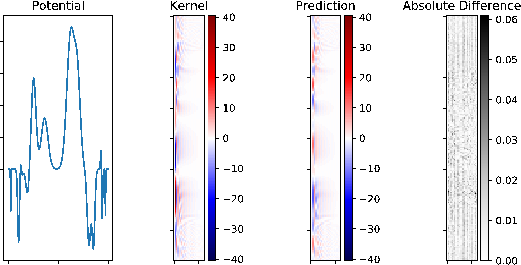}
\includegraphics[width=0.49\textwidth]{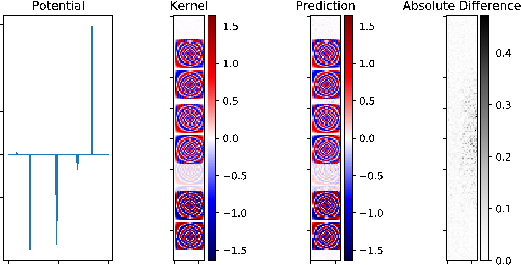}
\includegraphics[width=0.49\textwidth]{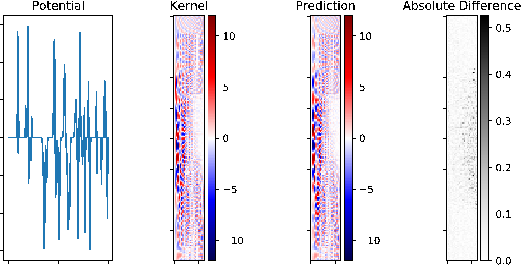}
\includegraphics[width=0.49\textwidth]{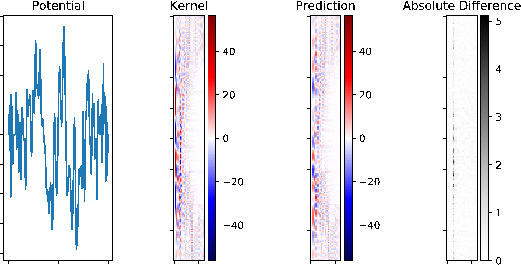}
\includegraphics[width=0.49\textwidth]{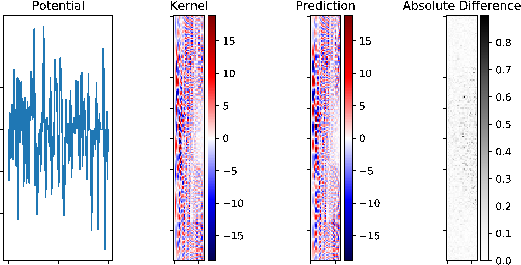}
\caption{The model is able to generalize to a large range of different potentials obtained by linear combinations of Gaussian functions.}
\label{fig:valid}
\end{figure}

\begin{figure}
\centering
\includegraphics[width=\textwidth]{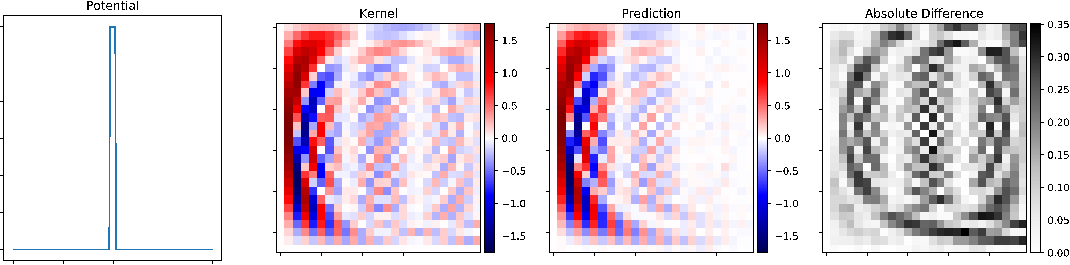}
\caption{The model is able to generalize to an abrupt potential barrier which is not part of the training set but which can be accurately approximated by narrow Gaussian functions (he kernel and its prediction are cropped to the region with non-zero values).}
\label{fig:test}
\end{figure}

\begin{table}
\centering
\begin{tabular}{ |c|c|c| } 
 \hline
 \# Potential & Avg. time (ms.) & Speedup \\ \hline
 1 & $1.69 \pm 0.4$ & 9x \\ \hline
 10 & $0.98 \pm 0.03$ & 15x \\ \hline
 25 & $0.89 \pm 0.03$ & 17x \\ \hline
 100 & $0.84 \pm 0.03$ & 18x \\ \hline
\end{tabular}
\caption{Computation time in milliseconds (ms) required to compute a complete Wigner kernel from a given potential.
By gathering several potentials together, the computation time per potential is amortized because of the efficiency of the GPU for matrix-matrix computations with high-dimensions. The average and standard deviation are computed over $10$ runs. The reference time for the computation time on the CPU is $15$ms.}
\label{table:time}
\end{table}

\section{Numerical validation}

In this section, for the sake of completeness,     we present a validation test which has
been already utilized in the previous two parts of this series.         In particular, we
simulate a representative one-dimensional quantum system made   of a Gaussian wave packet
moving against  a  potential barrier positioned at the center of a finite domain ($200$nm),
with width and height equal to $6$nm and $-0.3$eV, respectively.   The initial conditions
for this system consists of:
\begin{equation}
 f^0_W(x; M) = N e^{-\frac{(x-x_0)^2}{\sigma^2}}
 e^{-\frac{1}{\hbar^2}(M \Delta p - p_0)^2 \sigma^2}
\end{equation}
with $N$, $p_0$, $x_0$ and $\sigma$ the constant of normalization.  As usual, the initial
position, dispersion and wave number of the packet are equal to $68.5$nm,      $10$nm and
$6.28 \cdot 10^{-2}$nm$^{-1}$, respectively. It can be easily shown that this corresponds
to an initial wave packet energy smaller than the energy of the barrier.     Consequently,
one expects both reflection and tunneling effects appearing    during  the time-dependent
simulation.         Finally, absorbing boundary conditions are applied at the edge of the
simulation     domain. In spite of its simplicity, this numerical experiment represents a
solid validation test. Obviously, more complex situations could be simulated but would be
out of the scope of this paper.

Concerning the training process of the neural network,     as a starting point, a dataset
must be created.  One is readily obtained by, first, discretizing the position into $200$
entries and $20$ entries for the momentum. Then,          we generate $20,000$ couples of
potentials and their corresponding Wigner kernel,        and a final split into a $90:10$
between the training set and the validation set is applied randomly. In more details, the
potentials are randomly selected from a family of Gaussian bell shapes     with different
number of peaks, positioned randomly and with different dispersions.      To validate the
training process on such set,  we adopt the standard cross-validation technique where one
trains several ANNs on the training set  and  select  the best one according to the error
on the validation set.                                      Moreover, we use as the final
test set a single  potential, which belongs to a completely different family than the one
used during the training phase.    More precisely, we test our newly suggested  ANN on an
abrupt potential barrier. Consequently,          we make sure that the model captures the
underlying structure of the transformation $f$.      By observations on various numerical
experiments performed (see below),       we conclude that this procedure is sufficient to
stabilize the loss across different random splits.  Moreover, one should note that from a
couple consisting of a potential and its corresponding kernel,    one can extract several
training examples, one per column of the kernel.          Therefore, in order to help the
optimizer of the training process,    we build mini-batches by gathering the columns of a
kernel sharing the same potential.

\begin{figure*}[h!]
\centering
\includegraphics[width=0.45\linewidth]{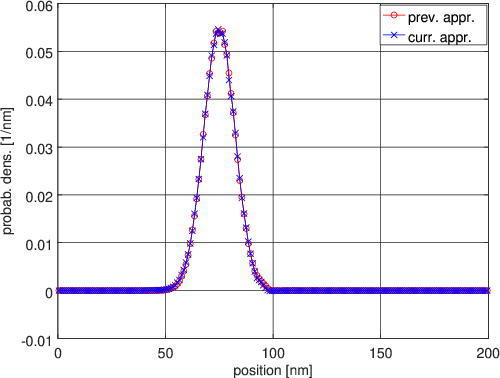}
\includegraphics[width=0.45\linewidth]{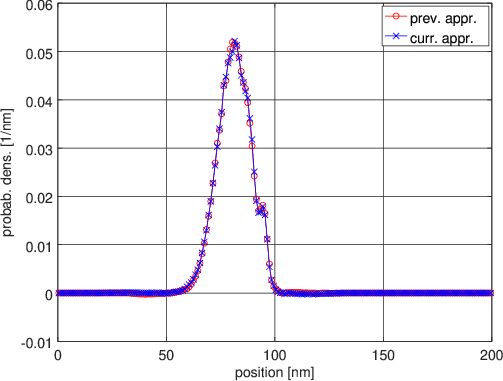}
\\
\includegraphics[width=0.45\linewidth]{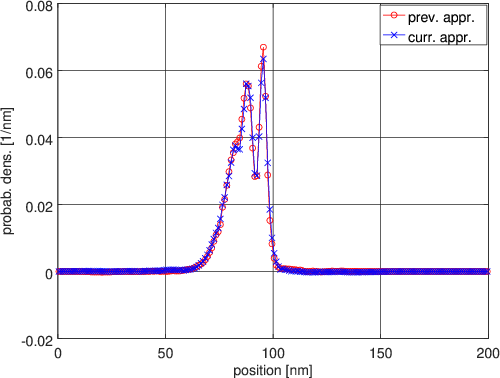}
\includegraphics[width=0.45\linewidth]{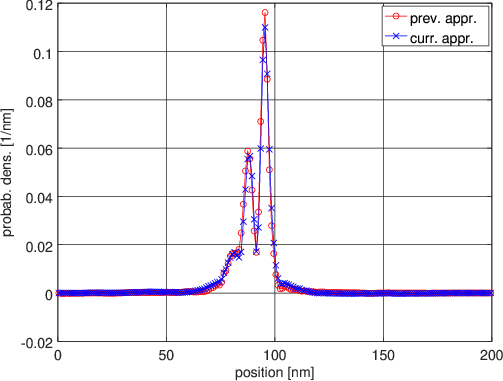}
\caption{Time-dependent evolution of a wave packet interacting with a potential barrier positioned at the center of the spatial domain,
at times $1$fs (top left), $2$fs (top right), $3$fs (bottom left) and $4$fs (bottom right) respectively,
and with two different approaches for the kernels corresponding to the case in \cite{PhysA-2017} (red circles) and the one suggested in this work (blue crosses).}
\label{probability}
\end{figure*}

The parameters of the model are optimized by means of the well-known ADAM method,   which
minimizes the Mean Square Error (MSE).    In particular, we use  a learning rate equal to
$10^{-4}$ which we manually decrease after $1500$ epochs by one order of magnitude.   The
size of the minibatch is equal to $10$ potentials, which is equivalent to $2000$ training
examples. The optimizer minimizes the MSE to $0.0011$ on the training set and $0.0041$ on
the validation set, as depicted in Fig. \ref{fig:learning}.  The use of ADAM is necessary
to obtain low errors in a reasonable time. \footnote{On a NVIDIA GTX 1080Ti, the training
time is less than $1$ day.}  Although in the presence of a model with $20$M parameters and no
regularization term at all, the overfitting proves to be unsignificant.       This can be
explained by the large number of training examples used to train the model. In fact, this
is confirmed by comparing the distributions of errors between        the training and the
validation sets. As we observe in Fig. \ref{fig:error} (right),            the normalized
distributions are hardly distinguishable.   Moreover, the distributions are highly skewed
towards zero,   and $95$\% of the examples have a mean absolute error smaller than $0.05$.
In Fig. \ref{fig:valid},   the potential input of the model is fixed and the index varies
in order to generate the whole kernel.      The kernels are transposed, so that the model
predicts each line of the kernel individually.         We observe that, once the model is
parametrized by a fixed potential,        it can generate accurately any position without
explicitly using the spatial information of the kernel. In Fig. \ref{fig:test},   we show
the prediction of the model for the barrier potential described above (the kernel and the
prediction are cropped to the region with non-zero values).   One observes that the model
is still able to provide accurate predictions even for low momenta,         but loses the
structure for high momenta with a maximum error around $0.35$.   While this error is high
relatively to the range of values for the kernel,               as previously observed in
\cite{PhysA-2017} and \cite{PhysA-2018},  the simulation of the physical system is robust
to this errors associated to high momenta. This is mainly due to two factors:  on the one
hand, the signed particle formulation is an intrinsically stochastic approach,  therefore
robust to noise/perturbations in the kernel, on the other hand, signed particles   rarely
explore the area corresponding to high momenta due to trivial energetic reasons.     As a
matter of fact,        these approximation errors do not have a significant impact on the
simulations as clearly shown in Fig. \ref{probability}.

\section{Conclusions}

In this work, the third of a series, we introduced a further technique   combining neural
networks and signed particles to achieve fast  and reliable time-dependent simulations of
quantum systems.    This newly suggested approach represents an important  generalization
over the previous techniques presented in \cite{PhysA-2017} and \cite{PhysA-2018}.     In
practice, we depicted a feedforward neural network,    i.e. a simple architecture able to
easily capture global interactions between the inputs,          and consisting of a layer
embedding the input, encoded as a one-hot vector, in a high-dimensional space   with five
times  more entries than the    number of positions, a layer transforming  this embedding
into a hidden representation concatenated to   the hidden representation of the potential
and two pairs of ReLU layers processing this vector followed by   a linear transformation
(see Fig. \ref{fig:model}).           Obviously, in this new context, a trade-off between
computational time and accuracy is clearly introduced.  Interestingly enough, in spite of
the simplicity of this model, we are actually able to achieve a speedup  about $18$ times
faster than the simplest finite difference quadrature     (see table \ref{table:time}). A  
representative validation test consisting of     a  wave  packet impinging on a potential
barrier has been performed which clearly shows      that the validity and accuracy of our
newly suggested method in practical situations.

Nowadays,  important experimental steps have been developed in the broad field of quantum
technologies,     such as quantum computing, quantum chemistry, nanotechnologies, etc. In
this promising context,   it is clear that our quantum simulation and design capabilities
are now starting to play a fundamental role which is going to grow in importance   in the
next future.       Consequently, solving modern technological problems related to quantum
mechanics is going to imply the adoption of dramatically different approaches.        The
authors of this paper believe that the technique suggested in this work        is a truly
promising candidate.

\bigskip

{\bf{Acknowledgments}}. One of the authors, JMS, would like to thank M.~Anti for her support, ethusiasm and encouragement.

\end{document}